\documentclass[conference]{IEEEtran}
\IEEEoverridecommandlockouts
\usepackage{hyperref}       
\usepackage{url}            
\usepackage{booktabs}       
\usepackage{amsfonts}       
\usepackage[cmex10]{amsmath}
\usepackage{amssymb}
\usepackage{amsthm}

\usepackage{algorithm}
\usepackage{algorithmicx}
\usepackage{algpseudocode}
\usepackage{graphicx}
\usepackage{subfigure}

\DeclareMathOperator*{\argmin}{arg\,min}
\DeclareMathOperator*{\argmax}{arg\,max}

\usepackage{cite}

\ifCLASSINFOpdf
\else
\fi
\usepackage{array}
\begin{document}
%
\title{Bayesian Inference on Matrix Manifolds for Linear Dimensionality Reduction}

\author{\IEEEauthorblockN{Andrew Holbrook}\thanks{Corresponding author Andrew Holbrook may be contacted at aholbroo@uci.edu.}
\and
\IEEEauthorblockN{Alexander Vandenberg-Rodes}\\
\IEEEauthorblockA{Department of Statistics\\
University of California, Irvine, California, USA}
\and
\IEEEauthorblockN{Babak Shahbaba}}


%


\maketitle

\begin{abstract}
We reframe linear dimensionality reduction as a problem of Bayesian inference on matrix manifolds. This natural paradigm extends the Bayesian framework to dimensionality reduction tasks in higher dimensions with simpler models at greater speeds. Here an orthogonal basis is treated as a single point on a manifold and is associated with a linear subspace on which observations vary maximally. Throughout this paper, we employ the Grassmann and Stiefel manifolds for various dimensionality reduction problems, explore the connection between the two manifolds, and use Hybrid Monte Carlo for posterior sampling on the Grassmannian for the first time. We delineate in which situations either manifold should be considered. Further, matrix manifold models are used to yield scientific insight in the context of cognitive neuroscience, and we conclude that our methods are suitable for basic inference as well as accurate prediction. All datasets and computer programs are publicly available at \url{http://www.ics.uci.edu/~babaks/Site/Codes.html}. 
\end{abstract}


%
\IEEEpeerreviewmaketitle

\section{Introduction}

The advent of Markov chain Monte Carlo in its many forms popularized Bayesian inference in the last decades of the 20$^{th}$ century. More recently, hybrid Monte Carlo (HMC)\cite{neal2011mcmc} has enabled efficient simulation from models with increasing numbers of parameters and deeper hierarchies.  Whereas HMC has extended fast Bayesian inference to higher dimensional models, high dimensional data analysis remains a lasting challenge for the statistical learning community. Linear dimensionality reduction is the most well established genre of dimensionality reduction tools. Famous examples from this toolkit include principal component analysis, canonical correlation analysis, and linear discriminant analyses (PCA, CCA, and LDA, respectively).

 PCA is a non-probabilistic linear dimensionality reduction technique in which the eigenvalue decomposition of the empirical covariance matrix is considered. Generative models include probabilistic PCA (PPCA) and factor analysis. Both of these methods model high dimensional data as generated by a multivariate Gaussian distribution with covariance the sum of a low-rank matrix and a diagonal matrix (restricted to a multiple of the identity under PPCA). Both have maximum likelihood (ML) as well as Bayesian implementations. The prevalent treatment of Bayesian PCA analysis is rather complicated: the columns of the ``loading matrix'' are modeled as independent multivariate Gaussian distributions each with its own variance hyper-prior \cite{bishop11bayesian}.
 
 In the following contribution we parsimoniously model the factor matrix as a single manifold valued parameter. Similar ideas in the context of factor analysis have been proposed in \cite{hoff2012model, chan2013invariant}, but we take a more general viewpoint. We unify a collection of models, including factor analysis and \emph{supervised PPCA} \cite{yu2006supervised}, along with exponential-family versions such as \emph{exponential} PCA~\cite{mohamed2009expca, li2013simple} and \emph{supervised} EPCA \cite{guo2009supervised, lu2012supervised, klami2012bayesian}. We distinguish between models parameterized by the Grassmannian and those defined on the Stiefel manifold.
 
 Our present approach is greatly informed by \cite{byrne2013geodesic}, in which the differential geometric framework of embedding geodesic Monte Carlo is established and implemented on the sphere and Stiefel manifold.  We also benefit from the matrix manifold optimization literature. Chief among these are \cite{edelman1998geometry, absil2004riemannian} and \cite{absil2009optimization}.
 Finally, follow the work of \cite{cunningham2015linear}, who argue for the relevance of matrix manifold optimization to the field of linear dimensionality reduction. The present results are largely extensible to non-probabilistic approaches discussed therein.

\section{Bayesian linear dimensionality reduction}
The most prevalent dimensionality reduction methods fall into the factor analysis framework. Such models specify the $N$ observed continuous data points $y_1,\dotsc, y_N\in \mathbb{R}^d$ as 
\begin{equation}\label{eq::FA}
    y_j = F z_j + \mu + \epsilon_j,
\end{equation}
where $z_j \in \mathbb R^k$ are the latent factors, $F$ is the $d\times k$ factor loading matrix, and $\epsilon_j$ are iid N$_d(0, \Psi)$ with $\Psi$ diagonal covariance matrix. Typically, the parameters $F,\mu, \Psi$ are optimized, either by EM or using closed form expressions available when $\Psi$ is restricted to be a multiple of the identity~\cite{tipping1999ppca}. If we place $N(0,I_k)$ priors on $z_j$, this latent factor is easily integrated out, leaving the sampling distribution, conditioned on $F,\mu,\Psi$, as 
\begin{equation}\label{eq::FA_marg}
y_j \sim N(\mu, F F^T + \Psi).
\end{equation}

This formulation assumes that the data lies close to an affine subspace spanned by the column vectors of $F$. There are, however, a wide continuum of subspaces that approximately span the data. Picking a single subspace can dramatically understate variation in the data and lead to over-fitting. One approach is to instead use the Bayesian framework to obtain a \emph{posterior} distribution over matrices $F$ which best explain the data. By thus integrating over high probability subspaces one may avoid over-fitting in a natural way. This was the approach taken in~\cite{mohamed2009expca}, in which the authors further generalized the factor analysis model to allow for observations $y_j$ from any exponential family distribution, allowing for principled dimension reduction for discrete data as well as continuous.

Despite its pleasing formulation, generating samples from the high-dimensional posterior distributions of $F$ proves to be especially difficult. Even using the very efficient hybrid Monte Carlo (see below), the model in ~\cite{mohamed2009expca} takes thousands of samples just to reach a high-probability region of the posterior.

The fundamental problem is that the model is highly over-parameterized on account of the rotational symmetry of the spherical Gaussian distribution. For any orthogonal $k\times k$ matrix $V$, we have 
\begin{equation}\label{eq::cancel}
(FV^T)(FV^T)^T = FF^T,
\end{equation}
 Furthermore, if the entries of $F$ are iid normal, then $FV^T$ has the same distribution as $F$. This causes significant problems for any MCMC sampler, as the resulting log-posterior $\log P(F, \mu, \Psi  | \mathbf y)$ is constant along $p(p-1)/2$-dimensional highly curved contours generated by the action of the orthogonal group. This high degree of curvature in the log-posterior can cause extremely low acceptance rates during sampling. 

One solution to the lack of identifiabiliy is to specify the factor loading matrix $F$ as upper triangular with positive entries on the diagonal \cite{lopes2004bayesian, carvalho2012high}. This specification has well-known problems due primarily to the ordering of the variables implied by the upper-triangular loading matrix \cite{chan2013invariant}. In what follows, we will explore an alternative approach and discuss its application in a more general dimensionality reduction framework.

\subsection{Reparameterizing with the Stiefel manifold}
Under the assumption of dimension reduction ($d > k$), the singular value decomposition (SVD) $F = U\Lambda V^T$ can be modified so that $U$ and $V$ are, respectively, $d\times k$ and $k\times k$ matrices with orthonormal columns, while $\Lambda$ is a $k\times k$ diagonal matrix with non-negative entries (the singular values) in decreasing order. Assuming the singular values are all distinct, $F$ is uniquely specified by $U, \Lambda$, and $V$. Now, recalling that $z\sim N(0, I_k)$ implies that $V^T z\sim N(0,I_k)$, we ignore the superfluous rotation by $V^T$ and reparameterize \eqref{eq::FA} into
\begin{equation}\label{eq::FArep}
y_j = U \Lambda z_j + \mu + \epsilon_j.
\end{equation}
The collection of $d\times k$ matrices with orthogonal columns, denoted by $\mathcal O_{d,k}$, is known as the real \emph{Stiefel manifold}, which is a (compact) Riemannian manifold of dimension $dk - \frac{k(k+1)}{2}$. Here, as in later models, we place a uniform prior distribution over $U\in \mathcal O_{d,k}$. We furthermore specify $\mu$ and $u_j$ to have independent mean-zero Normal priors, and the diagonal scale matrix $\Lambda$ has diffuse, positive valued priors on its entries.

We remark that similar models were considered in \cite{hoff2012model, chan2013invariant}, which model the SVD of the data, instead of the SVD of the factor matrix. In particular, \cite{hoff2012model} assumes that $Z\in \mathcal O_{N, k}$, where $Z^T = [z_1 \cdots z_N]$. The conditional distributions of each orthonormal column $P(U_j | U_{-j}, \Lambda, Z, Y)$ and $P(Z_j | Z_{-j}, \Lambda, U, Y)$ are shown to follow a von Mises-Fisher distribution, making the model amenable to Gibbs sampling. Relaxing this assumption to $Z_1,\dotsc, Z_N \sim N(0, I_k)$, as we do with \eqref{eq::FArep}, is not that different, at least a-priori. In most situations we have $N\gg k$, (even if the data dimension $d\approx N$) and the high-dimensional independent Gaussian random variables are orthogonal in prior expectation. 

This approach -- directly sampling the matrix parameter $U$ over the Stiefel manifold $\mathcal O_{d,k}$ -- generalizes to the models mentioned in the introduction, as we will show in the next section.
Our motivation is simple: \emph{when the relevant dimension reduction methods involve orthogonal projections, one should directly model the orthogonal projection, not an over-parameterized version.} In the case of factor analysis, i.e. when the entries of $\Psi$ are allowed to take on distinct values, it is often appropriate to parameterize by another manifold instead.

\subsection{Separated covariance model with the Grassmann manifold}

From \eqref{eq::FArep} it is clear that the PPCA/factor analysis model is a fully generative probability model for the observations $\mathbf y$, with the form
\begin{equation}
    y_j \sim N(\mu,  U\Lambda \Lambda U^T + \Psi),
\end{equation}
with $U\in \mathcal O_{d,k}$ and $\Lambda$ a diagonal $k\times k$ matrix.

The covariance matrix for $y_j$ can be expressed in a somewhat different fashion, as
\begin{equation}\label{eq::FAorthog}
y_j \sim N(\mu, \Phi (UU^T + \Psi) \Phi),
\end{equation}
where $\Phi$ is a diagonal scale matrix for the observations. This separation  strategy is similar to the approaches of \cite{liang1986longitudinal} in the context of generalized estimating equations and \cite{barnard2000modeling} in the context of Bayesian covariance modeling.   Model \eqref{eq::FAorthog} is distinct insofar as $UU^T + \Psi$ will almost surely never be a correlation matrix. The (non-negative) diagonal matrix $\Psi$ of residual variances is known as the \emph{uniquenesses} in the factor analysis literature \cite{johnson1992applied}. While in classical factor analysis the loading matrix $U$ can be an arbitrary $d$-by-$k$ matrix, we will here constrain $U$ to have orthogonal columns. That way, $UU^T$ is the unique projection matrix onto the subspace on which the standardized data
$\Phi^{-1} (y_j - \mu)$ approximately lie.

We remark that this model is still unidentifiable up to right rotations $U\mapsto UV^T$ with $V\in \mathcal O_{k,k}$, however, this can be solved by considering $U$ as an element of the quotient space $\mathcal O_{d,p} / \mathcal O_{k,k}$, which is invariant to right-rotations. This quotient space is known as the \emph{Grassmann} manifold $G_{d,k}$, which is in one-to-one correspondence with the set of $p$-dimensional subspaces of $\mathbb R^d$. In Section III we show how to perform inference on the Grassmann manifold while holding right rotations constant. Next, we will discuss a generalization of this approach to a broader class of models.

\subsection{Extension to exponential family PCA}
A significant limitation of standard PCA appears when one tries to apply it to binary, count, or categorical data. Taking a cue from generalized linear models, \cite{collins2001generalization} models each data point $y_{j,i}$ as coming from an exponential family distribution:
\begin{equation}
P(y | \theta) = h(y) \exp(y\theta - b(\theta)).
\end{equation}
Here the natural parameter $\theta$ is related to the mean $\mu = \mathbf E(y\ |\ \theta)$ through the canonical link function: $\theta = g(\mu)$, where $g^{-1}(\theta) = b'(\theta)$. Dimension reduction is then applied to the natural parameter $\theta$, which for many distributions of interest (e.g. Bernoulli or Poisson) can take any value on the real line, unlike mean $\mu$. 

For simplicity we describe the case of binary data. With $X_j\in \{0,1\}^d$, we have the canonical logit link function $g(p) = \log \frac{p}{1-p}$ and 
\begin{align}
    X_{j,i} &\sim Bernoulli(p_{j,i}), \quad \mbox{with} \label{eq::EPCA}\\
    p_{j,i} &= g^{-1}\left(\sum_{\ell=1}^k U_{i,\ell}\lambda_\ell z_{j,\ell} + \mu_i\right),\label{eq::EPCA2}
\end{align} 
where the $X_{j,i}$ are conditionally independent given the parameters $(U, \Lambda, \mathbf z, \mathbf \mu)$. We give parameters $U, \Lambda, \mathbf z, \mu$ the same priors as in \eqref{eq::FArep} and recall that $U\in \mathcal O_{d,k}$. 

In \cite{collins2001generalization}, $\Lambda=I_k$ and $\mu$ is set to zero, but in particular $U$ is an unconstrained $d\times k$ matrix, with all parameters learned via (penalized) maximum likelihood. The Bayesian versions in \cite{mohamed2009expca, li2013simple} have a slightly different parameterization, but $U$ is again not restricted to the Stiefel manifold.

\subsection{Extension to PCA regression and classification models}
Suppose now that \emph{paired} data $(X, y)$ is collected. Often the goal is to fit a joint model for the data, such that future $X$ data can be used to predict $y$ -- the supervised learning problem. Two classical linear dimensionality reduction methods for this case are partial least squares (PLS) and linear discriminant analysis (LDA). In the case that the $y_j$ are class labels, LDA finds a projection $M\in \mathcal O_{d,k}$ of the $X$ data such that the ratio of the between-class variance $\Sigma_B$ to within-class variance $\Sigma_W$ is maximized~\cite{cunningham2015linear}:
\begin{equation}
\argmax_{M\in \mathcal O_{d,k}} \frac{\operatorname{tr}(M^T \Sigma_B M)}{\operatorname{tr}(M^T \Sigma_W M)}.
\end{equation}
With continuous data $y$, on the other hand, partial least squares is concerned with finding orthogonal projections of $X$ and $y$ to a common latent subspace such that their covariance is maximized.

Here, we focus on classification problems, specifically, variants of LDA. Often the data $X_j$ are high-dimensional binary or count valued data, which motivates a generalized linear model framework as with exponential PCA, for simultaneously modeling $X$ and $y$. We consider the following model:
\begin{align}
    X_j &\sim p(x\ | \theta_j), \mbox{ an exponential family vector} \label{eq::jointst} \\
    \theta_j &= g_X^{-1}(U\Lambda z_j + \mu) \label{eq::jointX}\\
    y_j &\sim p(y\ | \eta_j), \mbox{ an exponential family r.v.}\\
    \eta_j &= g_y^{-1}(\beta^T z_j + \beta_0) \label{eq::jointy}
\end{align}
We keep the prior specification for $U, \Lambda, Z, \mu$ in \eqref{eq::jointX} as before -- only the $\beta$ coefficients are new.

Later, we will discuss an application of this model to count data coming from neural spike trains, where we specify $X_j$ as Poisson with canonical link function $g(x) = \log x$, and model the response variable $y$ -- the behavioral response from a finite set of possible outcomes -- using (multinomial) logistic regression.

An example of this type of model is supervised logistic PCA \cite{yu2006supervised}, which was applied to genomic data. The parameters there were learned by maximum likelihood, with $U$ restricted to the Stiefel manifold.
As we saw before with exponential PCA, existing Bayesian versions such as \cite{klami2012bayesian} do not restrict $U$ to have orthonormal columns.

In contrast to prior Bayesian treatments, we model $U$ directly as a random element on the Stiefel and Grassmann manifolds and, in order to do so, employ the embedding geodesic Monte Carlo of \cite{byrne2013geodesic}.  

\section{Bayesian inference and the geodesic Monte Carlo}


Given data $y_1, \dots, y_N \in \mathbb{R}^n$, it is often useful to specify a generative model in the form of a likelihood function, $p(y | q)$. This is the forward model. In the following we allow $q \in \mathcal{M}^m$ to be an m-dimensional manifold valued vector that parameterizes the likelihood. Endowing $q$ with prior distribution $p(q)$ renders the posterior distribution
\begin{equation}
\pi_\mathcal{H}(q)=p(q | y) = \frac{p(y|q)p(q)}{\int p(y|q)p(q) \mathcal{H}^m(dq)} \; .
\end{equation}

The integral is often referred to as the evidence and may be interpreted as the probability of observing data $y$ given the model.  Here the prior distribution is defined with respect to the Hausdorff measure
\begin{equation}
 \mathcal{H}^m(dq) = \sqrt{|G(q)|} \lambda^m(dq) \; .
\end{equation}
This measure is exactly the Lebesgue measure $\lambda^m$ scaled by metric based volume element $\sqrt{|G(q)|}$, where $G$ is the Riemannian metric on $\mathcal{M}$.  Let $\pi_\mathcal{H}(q)$ denote the posterior density with respect to the Hausdorff measure.

For most interesting models the evidence integral is intractable and high dimensional models do not lend themselves to numerical integration.   Non-quadrature sampling techniques such as importance sampling or even random walk MCMC are similarly cursed.  

\subsection{Hybrid Monte Carlo}
Hybrid Monte Carlo is an effective sampling tool for high dimensional models with thousands of parameters.  Riemannian manifold HMC \cite{girolami2011riemann} is an extension with connections to Newton's method.  Embedding geodesic Monte Carlo is a further extension and is the basis for HMC on the Grassmann and Stiefel manifolds.

HMC is often referred to as Hamiltonian Monte Carlo.  One builds a Hamiltonian system from the posterior distribution $\pi(q)$ and an augmenting Gaussian variable $p$.  The negative-log transform turns the probability distribution functions into a potential energy function $U(q)$ and corresponding kinetic term $K(p)$.  Thus $q$ and $p$ become the position and momentum of Hamiltonian function
\begin{eqnarray}\label{eq::hamEuclid}
H(q,p) &=& U(q) + K(p) \\ \nonumber
       &=& - \log \pi(q) +\frac{1}{2} p^Tp \; .
\end{eqnarray}
In order to draw samples from $\pi(q)$, the system is numerically advanced according to Hamilton's equations:
\begin{eqnarray}
\frac{dq}{dt} &=& \frac{\partial H}{\partial p} \\ \nonumber
\frac{dp}{dt} &=& -\frac{\partial H}{\partial q}  \; .
\end{eqnarray}  

 In the case that variable $q$ takes values on a non-Euclidean space, the above representations are insufficient as curvature is not taken into account. On the other hand, if certain facts about the manifold of interest are known, one may overcome this difficulty by isometrically embedding the manifold into Euclidean space. 

Let $d$ be the standard Euclidean metric and denote the metric preserving embedding $x: (\mathcal{M}, G) \rightarrow (\mathbb{R}^{n}, d)$. This map renders a similar Hamiltonian to (6) above:
\begin{equation}
H(x,v) = - \log \pi_\mathcal{H}(x) + \frac{1}{2} v^Tv\;.
\end{equation}

Instead of Gaussian momentum $p \in \mathbb{R}^m$, we augment by Gaussian velocity $v \in T_x\mathcal{M}$, the tangent space to the embedded manifold at $x$.   Note that we write $\pi_\mathcal{H}$ now since $q$ is manifold valued and $\pi_\mathcal{H}(x) = \pi_\mathcal{H}(u)$ since $\mathcal{H}^m$ is invariant under isometric embeddings.

One may forward integrate the corresponding Hamiltonian equations by splitting the Hamiltonian into potential and kinetic terms\cite{shahbaba2014split}. The solution to the potential term is given by $x(t)=x(0)$ and
\begin{eqnarray}
v(t) = v(0) + t\, \Pi_{T_x\mathcal{M}} \big(\nabla_x \log \pi_\mathcal{H} (x) |_{x=x(0)} \big) \; .
\end{eqnarray}

Here $\Pi_{T_x\mathcal{M}}$ is the orthogonal projection onto the the tangent space to the embedded manifold at $x$. For the Stiefel and Grassmann manifolds, this map is available in closed form.  The solution to the kinetic term is given by the unique geodesic (with respect to the Levi-Civita connection) starting at point $x(0)$ and with initial velocity $v(0)$.

 The embedding geodesic Monte Carlo algorithm is presented in Algorithm \ref{alg1}.  To implement HMC on an embedded manifold only three quantities require evaluation: the log posterior density $\log \pi_\mathcal{H}$ and its gradients; the orthogonal projection from the ambient space $\mathbb{R}^{n}$ onto tangent space $T_x\mathcal{M}$; and geodesic flow associated with initial velocity $v \in T_x\mathcal{M}$.  Since the metric $G$ only appears in the Jacobian term of the prior density, there is no need to compute the metric $G$ when a uniform prior distribution is available.  Both the Grassmann and Stiefel manifolds admit a uniform distribution. They also have analytic geodesic flows as well as closed form projections onto the tangent space at any point $x$.    Thus they are suitable candidates for embedding geodesic Monte Carlo.
 
\begin{algorithm}  
  \caption{Embedding geodesic Monte Carlo\cite{byrne2013geodesic}} \label{alg1} 
  \begin{algorithmic}[1]
   \State $v \sim N(0, I_n)$
   \State $v \gets  \Pi_{T_x\mathcal{M}}(v)$
   \State $h \gets \log \pi_\mathcal{H}(x) - \frac{1}{2} v^Tv$
   \State $x^* \gets x$
   \For{$\tau = 1,\dots, T$}
   	\State $v \gets v + \frac{\epsilon}{2} \nabla_{x^*} \log \pi_\mathcal{H} (x^*)$
	\State $v \gets  \Pi_{T_x\mathcal{M}}(v)$
	\State \parbox[t]{\dimexpr\linewidth-3em}{Progress $(x^*,v)$ along the geodesic flow defined by initial velocity $v$.\strut}
	\State $v \gets v + \frac{\epsilon}{2} \nabla_{x^*} \log \pi_\mathcal{H} (x^*)$
	\State $v \gets  \Pi_{T_x\mathcal{M}}(v)$ 
   \EndFor
   \State $h^* \gets \log \pi_\mathcal{H}(x^*) - \frac{1}{2} v^Tv$
    \State $u \sim U(0, 1)$
    \If{$u < \exp (h^* -h)$ }
    	\State $x \gets x^*$
    \EndIf
  \end{algorithmic}  
\end{algorithm}

\subsection{Two matrix manifolds} 

The Grassmann manifold, denoted $G_k(\mathbb{R}^d)$ or $G_{d,k}$, is the space of $k$-dimensional subspaces of $\mathbb{R}^d$.  On this manifold, each point is a linear subspace of $\mathbb{R}^d$ as well as an equivalence class of all $d$-by-$k$ matrices the columns of which span the subspace.  The Stiefel manifold $\mathcal{O}_{d,k}$ is the space of orthonormal matrices of height $d$ and width $k$. See \cite{chikuse2003special} for an overview of the two manifolds and classical statistical inference. Both manifolds are smooth and compact, and both have been used to great success in non-Bayesian dimensionality reduction \cite{cunningham2015linear}.

Due to their compactness, both manifolds admit a uniform distribution with respect to the Hausdorff measure. This fact simplifies geodesic Monte Carlo since the uniform measure is constant and cancels in the accept-reject step. \cite{byrne2013geodesic} provides formulas for projection and co-geodesic flow of the Stiefel manifold, but (to the best of our knowledge) HMC has never been performed on the Grassmannian. We provide the necessary tools to do so.  

\subsection{Projection and flow on the Grassmann manifold}
For any point $X \in G_{d,k}$, the tangent space to $G_{d,k}$ at $X$  consists of the  rank $(d-k)$ subspace orthogonal to $X$.  That is, if we let $X_1$ be an orthonormal class representative with columns spanning $X$ (i.e., $[X_1] =X$), then the projection onto the tangent space at $X$ is given by the simple formula
\begin{eqnarray}\label{eq::grass_proj}
\Pi_{T_XG_k(\mathbb{R}^n)} &=& \big( I_n - X_1(X_1^TX_1)^{-1} X_1^T \big) \\ \nonumber
&=&  \big( I_n - X_1 X_1^T \big) \; .
\end{eqnarray}

From here on we conflate point $X \in G_{d,k}$ with any orthonormal matrix $X_1$ satisfying $[X_1]=X$ and vice-versa.  Given an orthonormal representative $X(0) \in G_{d,k}$, any vector $\dot{X}(0) \in T_X G_{d,k}$ determines a unique geodesic path with respect to which $\dot{X}(0)$ acts as initial velocity.  In order to compute this path, we require the singular value decomposition $\dot{X}(0) = U \Sigma V^T$.  Once we have this decomposition, the geodesic path is given by
\begin{eqnarray} \label{eq::grass_geo}
\Big(X(t), \dot{X}(t)\Big) =& \Big(X(0) V,X(0) \Big) \times \\ \nonumber
&\left( \begin{array}{cc}
\cos \Sigma t & - \sin \Sigma t  \\
\sin \Sigma t & \cos \Sigma t  \end{array} \right) \left( \begin{array}{c}
V^T \\
\Sigma V^T \end{array} \right)\; .
\end{eqnarray}
 
We attain \eqref{eq::grass_geo} by differentiating formula (2.65) of \cite{edelman1998geometry}. This formula includes the familiar rotation matrix applied element-wise to the elements of $\Sigma$.  It is easy to see that $X(t)^TX(t)=I_k$. That is, the geodesic formula advances orthonormal class representative to orthonormal class representative. In \eqref{eq::grass_geo} $V$ acts as a random right rotation on $X(0)$, but the geodesic remains well-defined even when $V$ is fixed to any orthogonal matrix \cite{edelman1998geometry}.  Due to the fact that $X(t)$ is orthonormal, allowing $V$ to vary causes both Grassmann and Stiefel geodesic Monte Carlo to perform similarly. When $V$ is fixed, however, geodesic Monte Carlo is performed directly over subspaces of $\mathbb{R}^d$ and mere changes of basis are never considered.

 In order to implement geodesic Monte Carlo on the Grassmannian, one need only select any orthonormal $d$-by-$k$ matrix, $X_0$, then advance Algorithm 1 by implementing  \eqref{eq::grass_proj} in algorithm lines 2 and 7, and implementing \eqref{eq::grass_geo} in line 8.

\section{Convergence and posterior summaries using the projection Frobenius metric}

As model dimensionality increases, it becomes increasingly difficult to calibrate parameters and assess model fit to data.   This is particularly true when a model is built for learning higher moments and latent factors.   When assessing model effectiveness with simulated data, it is important to have a measure of closeness to truth, while for real data it is important to understand the uncertainty of parameter estimates, whether through confidence intervals or posterior distributions. 

In the context of the above factor analysis type models, one often has a collection of samples $U_1,\dotsc, U_n$ of the factor loading matrices, each of which provides a different subspace $\operatorname{Range}(U_j)$ on which the data is assumed to approximately lie. Crucially, we are interested in understanding the variability in these \emph{subspaces}, instead of the variability of the matrix elements. This is because for any $V\in \mathcal O_{k,k}$ the matrix $U V^T$ describes the same subspace -- multiplying by $V^T$ merely rotates the basis vectors within it. 

Hence the Grassmann manifold $G_{d,k}$ (being the space of linear subspaces) is the very space within which we would like to characterize variability. In the following we explore this manifold's projection Frobenius (pF) metric, which is easily used for diagnosing convergence of the MCMC chain, as well as for assessing the variability of the posterior distribution of $U$. In addition to being high dimensional, the posterior distribution of matrix representative $U$ is typically multi-modal (if not unidentifiable) on account of the symmetry under rotations by $V$. Therefore traceplots of its entries are hard to diagnose for convergence.  On the other hand, metrics on the Grassmann manifold are agnostic to such rotations. Thus the pF distance between the samples and a reference point proves much more informative.

\subsection{Projection Frobenius distance on the Grassmann manifold}

There are a variety of metrics that have been defined on $G_{d,k}$, all of which are easily calculated in terms of the \emph{principle angles} between $k$-dimensional linear subspaces $X$ and $Y$ in $\mathbb{R}^d$. See \cite{edelman1998geometry} for short discussion. These angles are defined with respect to the SVD of the matrix representatives: if we have $X^TY = U \cos \Theta V^T$ with $U, V \in \mathcal O_{k,k}$ and $\Theta$ a non-negative diagonal matrix, then the principle angles are the singular values $\theta_1,\dotsc, \theta_k \in [0,\frac{\pi}{2}]$  \cite{edelman1998geometry}.

Letting $\lVert\cdot\rVert_F$ denote the Frobenius norm, the pF distance between $X$ and $Y$ is defined as
 \begin{eqnarray}\label{eq::dchord}
 d_{pF}(X,Y) = \frac{1}{2} \lVert XX^T-YY^T\rVert_F =\sqrt{\sum_{j=1}^k \sin^2\theta_j} \; ,
 \end{eqnarray}
while the closely related \emph{geodesic} distance is
 \begin{equation}
 d_g(X,Y) = \sqrt{\sum_{j=1}^k \theta_j^2} \; .
 \end{equation}
 
Note that $d_{pF}(\cdot,\cdot)$ has maximal distance $\sqrt k$.

\subsection{The projection Frobenius mean}
Given a collection of samples $U^{(1)},\dotsc, U^{(N_s)}$ from the Grassmann manifold, we would like to assess traceplots of the distances between these samples and a reference point -- typically the posterior mean. Since the Grassmann manifold is not a vector space, we use the idea of a \emph{Karcher} mean, which is a point $U^{(0)}$ minimizing the average distance to the samples:
\begin{equation}
U^{(0)} = \argmin_{U\in G_{d,k}} \sum_{j=1}^{N_s} d(U, U^{(j)}). 
\end{equation}
Under the pF metric we will refer to the Karcher mean as the \emph{pF mean}. The pF mean has a simple closed form formula requiring only a single SVD, unlike the Karcher mean using the geodesic metric \cite{harandi2015extrinsic}.

\begin{figure}
  \centering
\centerline{\includegraphics[width=\linewidth]{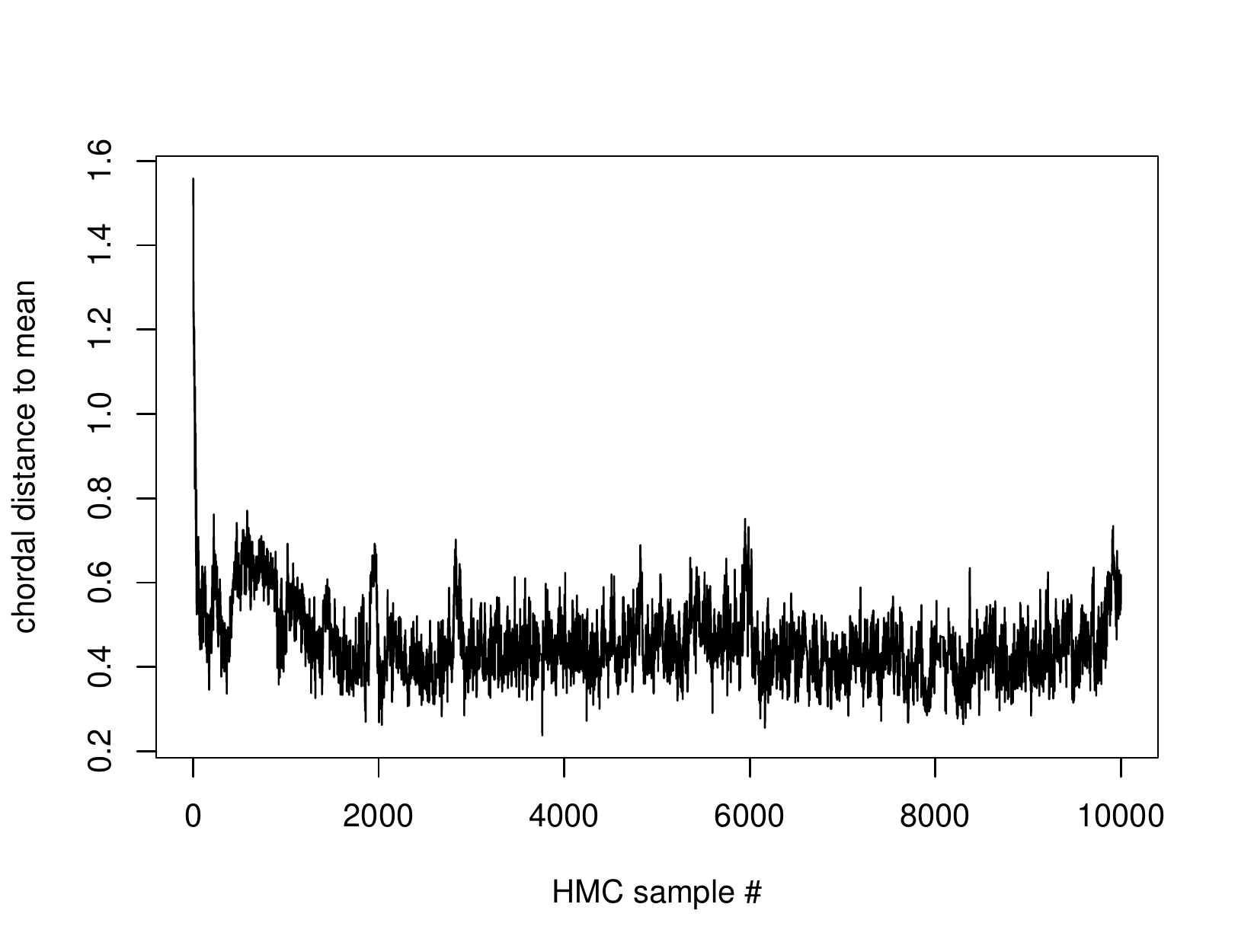}}
\caption{Traceplot of pF distance between the $U$ samples and their pF mean for the synthetic data example.}
\label{fig::dchordal}
\end{figure}

For illustration, in Figure~\ref{fig::dchordal} we show samples $U^{(s)}\in G_{16,3}$ of the orthogonal loading matrix for the exponential-family PCA experiment discussed in the following section. We first compute the chordal mean $U^{(0)}$ of the latter samples $U^{(5001)},\dotsc, U^{(10000)}$, and then plot the pF distance \eqref{eq::dchord} between $U^{(0)}$ and all the samples $U^{(s)}$, $s=1,\dotsc, 10000$.  This technique is interpretable even if samples are drawn using Stiefel geodesic Monte Carlo.  One simply identifies the point on the Stiefel manifold, an orthogonal matrix, with the subspace its columns span.  

Note that after about sample 30, all loading matrix samples lie within a pF distance $0.8$ of the pF mean. We caution against low-dimensional intuition: despite the pF distance between points on $G_{16, 3}$ being at most $\sqrt 3$, for $r<1$ the metric ball
\begin{equation}
B(r) = \{U \in G_{d,k}\ :\ d_{pF}(U, U_0) < r\}
\end{equation}
has exponentially small volume (in $d$ and $k$) under the uniform measure \cite{dai2008quantization}. E.g., in our example $|B(0.8)|\approx 10^{-11}$.

\section{Results}

\subsection{Synthetic data}\label{sec::EPCA}
\begin{figure*}
    \centering
    \includegraphics[width=\textwidth]{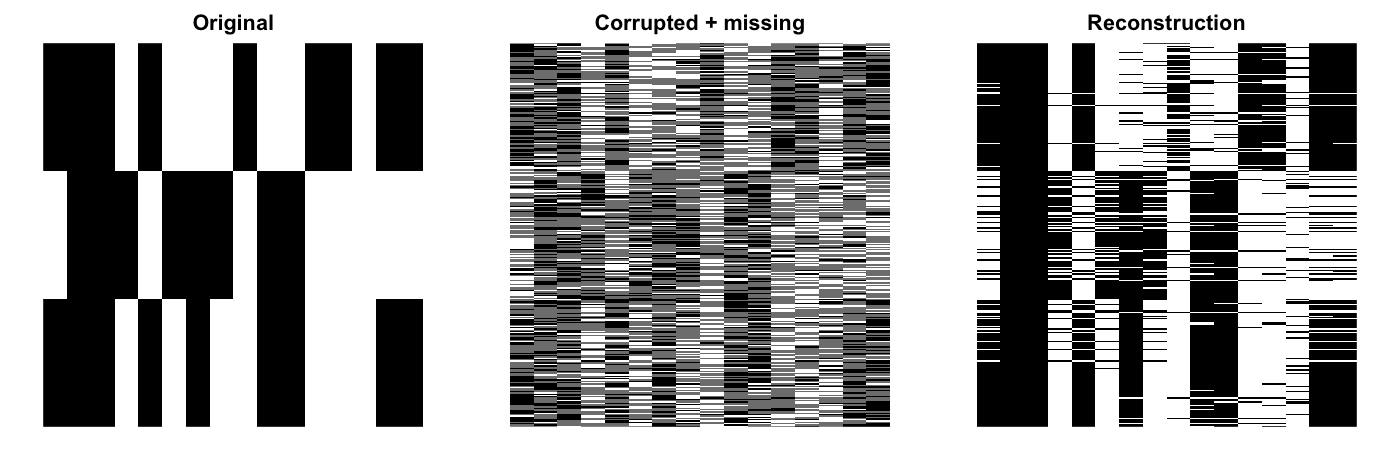}
   \caption{From left to right: Original samples of bit vectors; After randomly corrupting 20\% and setting 50\% as missing (grey pixels); Reconstruction averaging over 401 Stiefel HMC samples}.
    \label{fig::bitvector}
\end{figure*}

\begin{figure}
    \centering
    \includegraphics[width=0.95\linewidth]{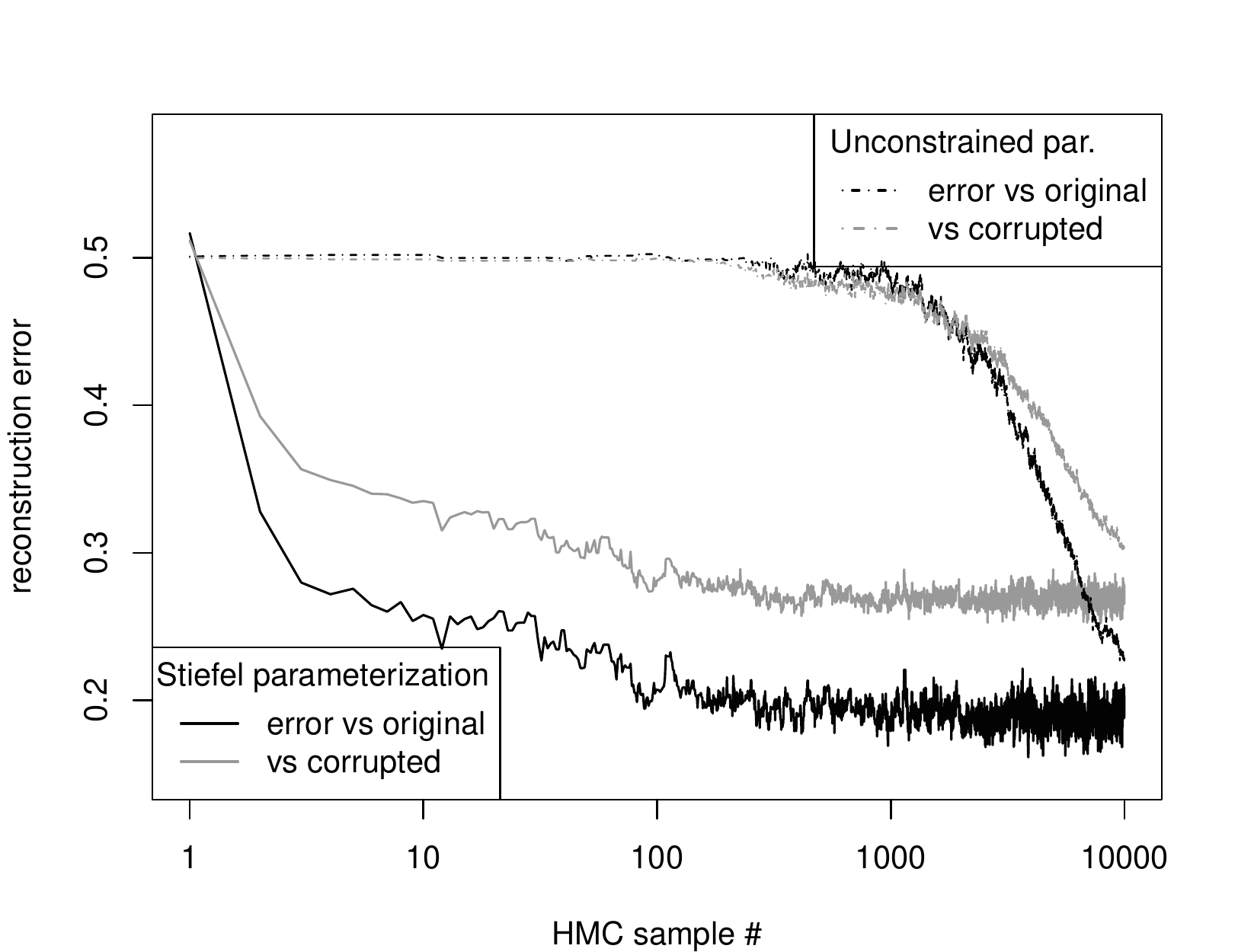}
    \caption{Traces of reconstruction error under HMC sampling of the exponential PCA model, comparing the Stiefel manifold parameterization to the unconstrained model of \cite{mohamed2009expca}.}
    \label{fig::bitresults}
\end{figure}

We illustrate here the difference between an over-parameterized factor analysis model and the model \eqref{eq::EPCA}, \eqref{eq::EPCA2} with loading matrix constrained to lie on the Stiefel manifold.
Recall the simulated data from~\cite{mohamed2009expca}: three 16-bit strings were chosen at random, and each repeated 200 times to form 600 binary vectors. Each bit was flipped independently with probability 20\% (10\% in \cite{mohamed2009expca}), giving a corrupted set of vectors $y_1,\dotsc,y_{600}$. We then randomly sample half (4800/9600) of the entries of $\mathbf y$ and set them as missing data. (see Figure \ref{fig::bitvector}). Next we fit this data to the binary logistic regression formulation of~\eqref{eq::EPCA}, \eqref{eq::EPCA2} with just three latent dimensions, and compare with the model of \cite{mohamed2009expca} with unconstrained loading matrix $U$.
The first question is how quickly the models fit the corrupted and half-missing data; second, whether the low-rank assumption allows for accurate reconstruction of the original data, despite only 50\% of the corrupted data being available. The results are shown in Figure~\ref{fig::bitresults}. Note that the HMC sampler for the Stiefel manifold model immediately reconstructs most of the missing and corrupted data: after a mere 100 Stiefel HMC samples, the \emph{per sample} error rate has reached an equilibrium of about 20\%. The unconstrained model needs over 10,000 samples to attain a similar accuracy. Both HMC samplers are tuned to achieve an acceptance rate of 60-80\%, and with $L=80$ steps per sample. By averaging over just samples 100,101,...,500 the Stiefel model obtains an error rate of 10.7\%. The right most panel of Figure~\ref{fig::bitvector} shows this Bayesian reconstruction. Averaging over the remaining 9,500 samples only drops the error rate to 9.4\%.

\subsection{Real data}
Next, we apply our models to three separate data sets: the first is the 18-dimensional \textit{Tobamovirus} data from \cite{tipping1999ppca, ripley2007pattern}; the second is the $52-$dimensional metabolite dataset of \cite{scholz2005non}; and the third is $53$-dimensional neural spike train data from an experiment conducted at (university name, here). We use the first data set to compare predictive accuracy of the Grassmann-manifold factor analysis model \eqref{eq::FAorthog} against that of its maximum likelihood counterpart. For the second dataset we measure imputation performance of the PPCA model \eqref{eq::FA} with missing data, specifically comparing geodesic Monte Carlo to the Variational Bayes method. For the neural firing data we employ the supervised Poisson-Logistic model of \eqref{eq::jointst} -- \eqref{eq::jointy}, and compare to other popular supervised classification algorithms.  Gradients and log-probabilities are computed using \cite{gelman2014rstan}. Computation is also performed using \cite{hoff2012rstiefel, ggplot2}.

\subsubsection{Tobamovirus data}

\begin{figure}[t]
  \centering
   \includegraphics[width=\linewidth]{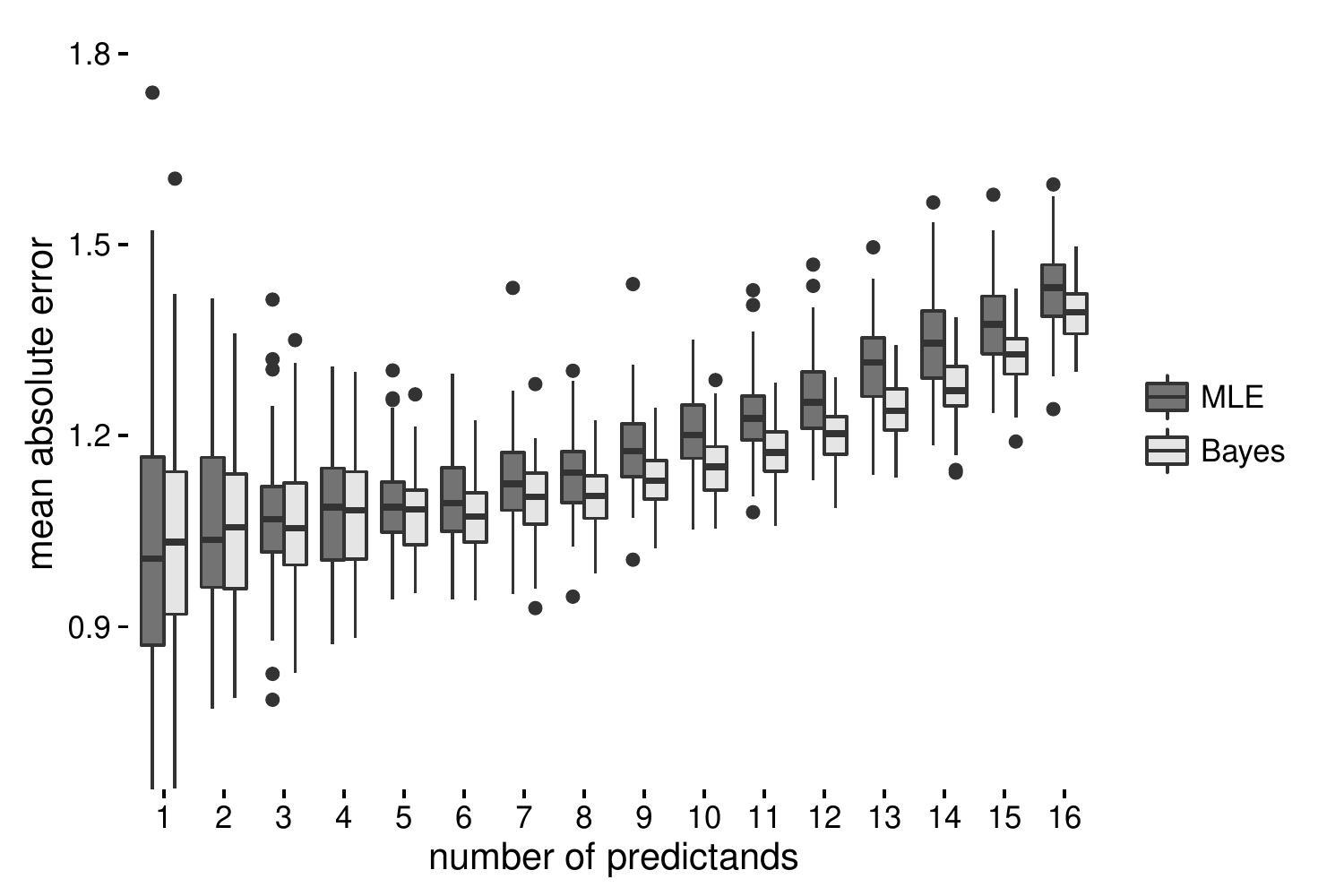}
\caption{Tobamovirus data, comparing the Grassmann manifold Bayesian model to standard MLE model. The 18 elements of the hold-out vector are assigned to be predictors and predictands, with the displayed prediction error an average over 100 random assignments. As the number of predictands increases (and number of predicting elements decreases) element-wise errors increase, and the Bayesian implementation outperforms amid greater uncertainty.}
\label{fig::Tobama}
\end{figure}

We compare the Bayesian-Grassmannian factor analysis model \eqref{eq::FAorthog} to its maximum likelihood counterpart with respect to predictive accuracy. In order to do so, we implemented a modified leave-one-procedure (LOO) on the Tobamovirus dataset.  The Tobamovirus data features 38 observations of vectors in $\mathbb{R}^{18}$.  For 38 iterations of LOO we trained both the Bayesian and the MLE models on 37 observations.  Next, we randomly divided the hold-out observation into elements to predict (predictands) and elements upon which to condition (predictors).  We then computed the Gaussian conditional mean for the predictands, given the predictors.  We chose the mean absolute ($L^1$) distance of predictand from conditional mean as prediction criterion.  Among other things, the conditional mean is a function of covariance matrix $\Sigma$.  Therefore, if one model has better prediction with respect to conditional mean, we may conclude that its low-dimensional representation $\Sigma = U\Lambda U^T + \sigma^2 I_d$ (in the case of PPCA) is superior. In addition to comparing mean LOO error, we show the methods' sensitivity to sparse predictors. Figure \ref{fig::Tobama} shows the results, with the Bayesian-Grassmannian implementation having lower prediction error as the number of predictors decreases. The Bayesian model was initialized at the MLE estimates, with $U$ orthogonalized. We used a scant 200 samples for each LOO interation (without thinning), as traceplots of the pF distance \eqref{eq::dchord} of samples $U$ to the MLE showed extremely low autocorrelation.

\subsubsection{Metabolite data}
\begin{figure}
\centering
\centerline{\includegraphics[width=0.95\linewidth]{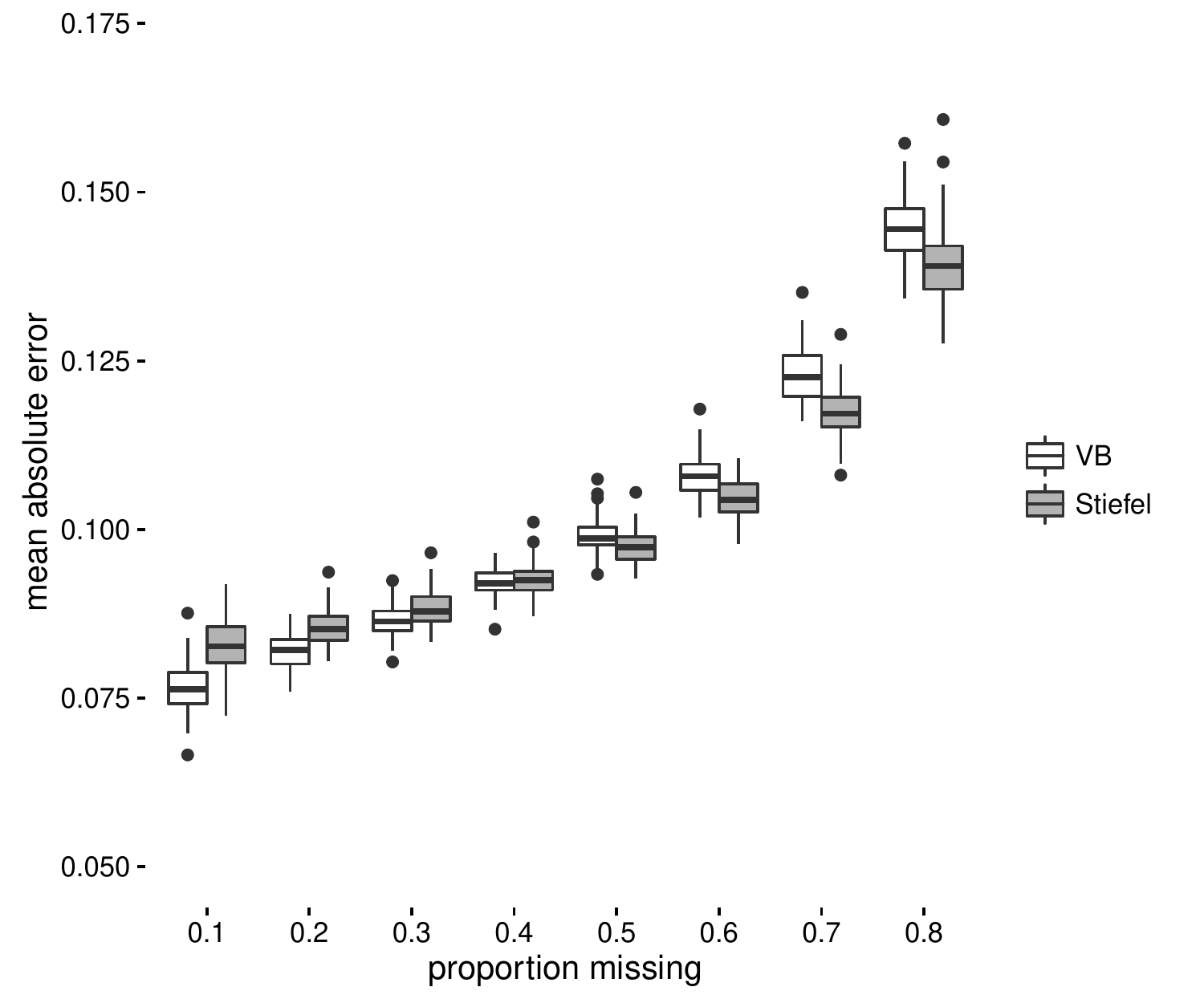}}
\caption{Imputation performance on metabolite dataset, comparing Variational Bayes to Stiefel HMC. As uncertainty increases, the fully Bayesian treatment excels. Both methods are presented at optimal number of latent factors.  For Variational Bayes, 52, for the Stiefel HMC model, there are diminishing returns after 7 latent dimensions}
\label{fig::metabolite}
\end{figure}

We compare the performance of the Bayesian-Stiefel PPCA model \eqref{eq::FA} to Variational Bayes \cite{bishop1999variational, oba2003bayesian}, when used for infilling missing data. We consider the \emph{metabolite} dataset of \cite{scholz2005non}. The dataset consists of 154 vectors of length 52, which are log-ratios of the concentrations over time (compared to a baseline) of 154 metabolites in a cold-stress experiment. We randomly assigned a percentage of the datapoints (independently at random) to be missing, and calculated the mean absolute reconstruction error for the two models. We varied the percentage of missing data to be 10\%, 20\%,$\dotsc$, 80\%, and plot the results of 100 trials for each percentage in Figure \ref{fig::metabolite}. Both methods perform better as one increases the number of latent dimensions (due to the built-in Automatic Relevance Determination through the priors on the scales $\Lambda$). For the Variational Bayes method we choose the best performance by fixing the number of latent dimensions to the maximum of 52.  For the Stiefel sampling model we made do with just 7 latent dimensions. Despite this handicap, the Stiefel sampling model does significantly better than Variational Bayes when 50\% or more of the data is missing. This result accords with that of Tobamovirus data: as uncertainty increases, fully Bayesian treatments excel.

\subsubsection{Neural spike data}

The neural spike data comes from a non-spatial sequential memory experiment on rodents \cite{allen2016nonspatial}.  Neural activity was recorded in the hippocampi of 6 rats, who had previously been trained on a particular "correct" sequence $(A,B,C,D,E)$ of odors. 
Each trial involves the rat smelling one of the five odors through a port. The rat signals whether the odor is \emph{in sequence} (InSeq) or \emph{out of sequence} (OutSeq). It does this by choosing to withdraw its nose from the port  either after or before one second, signalling InSeq or OutSeq respectively. 

Note that with the "correct" sequence $(A,B,C,D,E)$, each presented odor is InSeq. Whereas with the sequence $(A,B,C,C)$ the first three odors are InSeq while the last odor (the repeated odor "C") is OutSeq. About 88\% of the trials were in sequence. In this section we only look at data from a single session featuring rat Super Chris.  The session consists of 249 trials lasting anywhere from 0.48 to 1.74 seconds each. The data features spike counts from 53 neurons and a binary indicator for whether the present odor is InSeq (1) or OutSeq (0). In order to minimize differences in motor neuron activity across trials, the spike counts for each trial are the total number of spikes in the 0.4 second interval immediately preceding port withdrawal. We are interested in a supervised learning problem: can we decode the rat's response (InSeq vs OutSeq) from the spike data alone?

We use the Poisson-Logistic joint model \eqref{eq::jointst} -- \eqref{eq::jointy}. Spike count vectors $X_t\in \mathbb{ R}^{53}$ are modeled as conditionally independent Poisson, with log rate vector $U\Lambda z_t + \mu$ assumed to have rank $5$ -- that is, the latent factors $z_t$ are in $\mathbb{R}^5$. The binary in-sequence, out-of-sequence variable $y_t$ is modeled using logistic regression on the latent factors $z_t$. Hence the latent factors play two roles: they explain the majority of variation in spike counts \textit{and} they predict for sequence status.

\begin{figure}
    \centering
    \includegraphics[width=0.95\linewidth]{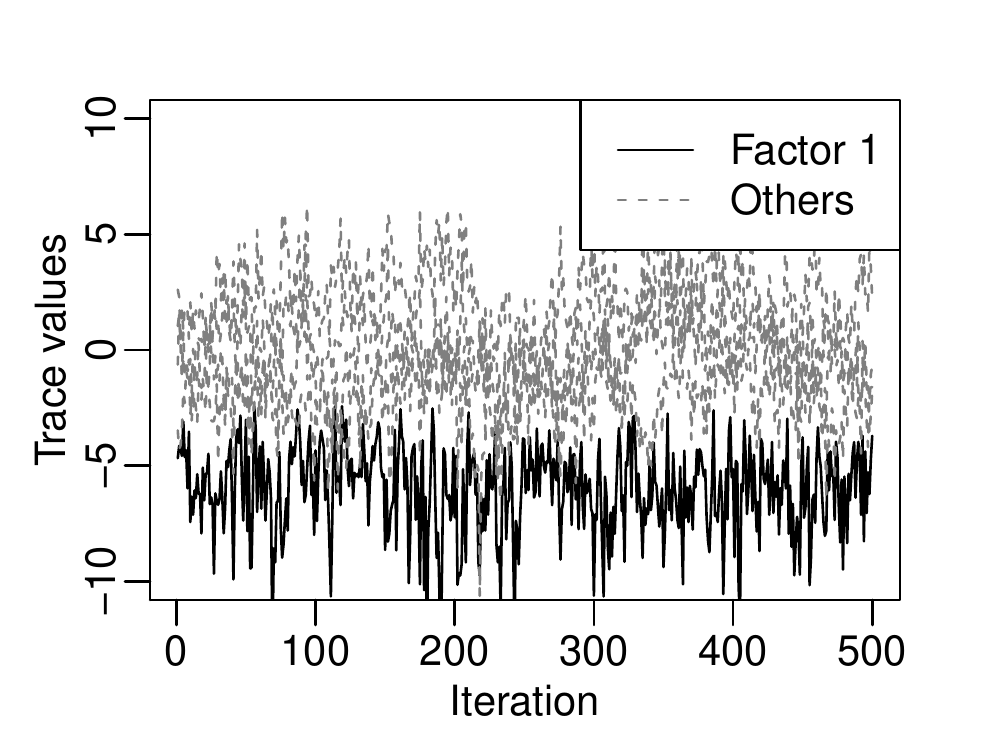}
   \caption{The posterior trace plots of logistic regression parameters associated with the first latent factor (black) and remaining minor factors. Created using 500 geodesic Monte Carlo samples}
    \label{fig::rat_trace}
\end{figure}

The logistic regression parameters $\beta$ are of particular interest.  These coefficients directly relate patterns in spike count to whether a sequence is correctly ordered.  Their distributions may support the scientific hypothesis that the rat hippocampus is a place where sequential learning is performed.  Here \textit{learning} is meant to suggest a global phenomenon, one involving relationships between individual neurons and groups thereof.  Figure \eqref{fig::rat_trace} affirms the hypothesis in a specific sense: if a coefficient has a significantly non-zero posterior sample,  then intensity of the relevant factor corresponds to the increased or decreased odds of sequential correctness.

Figure \eqref{fig::rat_trace} features 500 draws from the posterior distributions of the logistic coefficients associated with the five latent factors. We simulated 10,000 samples, discarded the first half, and thinned nine of every ten draws. The first parameter has a distinctly non-zero posterior, while the rest do not. We infer that the first latent factor has a statistically significant association with sequential correctness.  Moreover, the strictly negative posterior suggests that this association is in fact negative.  

Besides providing interpretable regression coefficients,  the joint model outperforms competitors with respect to prediction accuracy. Table 1 shows prediction error rates for a number of methods under (0-1) loss.  All methods use spike count data or a reduction thereof to predict sequential correctness.  As 88\% of odors are presented in-sequence, uniformly predicting in-sequence earns the low error rate of $0.12$. PLS-DA predictions are made by first performing PLS-DA \cite{barker2003partial} for dimensionality reduction then running one of the respective prediction methods on the reduced data.       Only the Bayesian joint model is able to correctly predict a significant fraction of out-of-sequence odors, achieving an error rate below $0.08$.

\begin{table}
\centering
\caption{10-Fold cross-validation error}
\begin{tabular}{lc}
  \hline
Method  & 0-1 Error\\ 
 \hline
 Bayesian joint model &  0.076     \\
 Random forest &  0.103  \\
 PLS-DA SVM & 0.111 \\
 PLS-DA 5NN & 0.119 \\
 PLS-DA LDA & 0.124 \\
   \hline
\end{tabular}
\end{table}

\section{Discussion}

We used geodesic Monte Carlo on the Stiefel and Grassmann manifolds to extend Bayesian analysis to linear dimensionality reduction models for high-dimensional data. By reparameterizing earlier versions of (exponential-family) PPCA and factor analysis, we demonstrated dramatically more efficient sampling. We showed how to perform geodesic Monte Carlo on the Grassmann manifold and demonstrated use of the Grassmannian pF distance for diagnosing convergence of both Stiefel and Grassmann manifold-valued parameters. We compared our manifold parameterized models to maximum likelihood counterparts and state-of-the-art Bayesian implementations, with favorable results. And in the the context of neural spike trains we demonstrated how the manifold parameterization allows for efficient Bayesian analysis of more complicated supervised dimensionality reduction tasks, resulting in superior prediction accuracy on held out data.

The above applications are in no way comprehensive. Indeed one may use Bayesian inference on the Stiefel and Grassmann manifolds to probabilize many of the methods found in \cite{cunningham2015linear}.  These new implementations will not necessarily resemble past iterations of probabilistic linear dimensionality reduction. We have shown that the geodesic Monte Carlo does not restrict dimensionality reduction to simple models but allows for inclusion into broader joint or graphical models.  In turn, Bayesian dimensionality reduction becomes a tool for scientific inference as well as prediction.



%
\bibliographystyle{plain}
\bibliography{GHMCrefs}

\end{document}